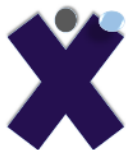

# Application of the Pythagorean Expected Wins Percentage and Cross-Validation Methods in Estimating Team Quality

Christopher Boudreaux[1], Justin Ehrlich[2], Shankar Ghimire[3,*], and Shane Sanders[4]

[1] Department of Economics, Florida Atlantic University, Boca Raton, FL, [2] Department of Sport Management, Syracuse University, Syracuse, NY, [3] School of Accounting, Finance, Economics, and Decision Sciences, Western Illinois University, Macomb, IL, [4] Department of Sport Management, Syracuse University, Syracuse, NY

*Corresponding Author E-mail:
sp-ghimire@wiu.edu

**Abstract**
The Pythagorean Expected Wins Percentage Model was developed by Bill James (1980) to estimate a baseball team's expected wins percentage (as distinct from the team's actual wins percentage) over the course of a season. As such, the model can be used to assess how lucky or unfortunate a team was over the course of a season (actual wins – expected wins). From a sports analytics perspective, such information is valuable in that it is important to understand how reproducible a given result may be in the next time period. In contest-theoretic (game-theoretic) parlance, James' original model represents a (restricted) Tullock contest success function (CSF). We transform, estimate, and compare James' original model and two alternative models from contest theory—the serial and difference-form CSFs—using MLB team win data (2003-2015) and perform a cross-validation exercise to test the accuracy of the alternative models. The serial CSF estimator dramatically improves wins estimation (reduces root mean squared error) compared to James' original model, an optimized version of James' model, or an optimized difference-form model. We conclude that the serial CSF model of wins estimation substantially improves estimates of team quality, on average. The work provides a real-world test of alternative contest forms.

**Keywords**: expected wins estimation, expected output, quality estimation, labor contest, sports analytics, cross-validation

## 1 The Pythagorean Expected Wins Percentage Model: Origins and Mathematical Applications

The Pythagorean Expected Wins Model was developed by Bill James (1980) to estimate a baseball team's expected wins (as distinct from the team's actual wins) over the course of a season. As such, the model can be used to assess how lucky or unfortunate a team was over the course of a season (actual wins – expected wins). Such knowledge is valuable, as future outcomes depend upon expected rather than realized outcomes. From a managerial perspective, it is important to understand how reproducible a given result may be in the next time period. Anthony et al. (2011), Drayer et al. (2012), and Mungeon and Winfree (2012) each show that team quality remains a primary determinant of gate and television demand for major league sports in North America. Moreover, a process-oriented organization may wish to reward employees based upon expected rather than realized outcome.

The Pythagorean model is, essentially, an application of mathematics-related models of expected wins percentage (see, e.g., Hadley et al. 2000). The Pythagorean model utilizes an economic concept (contest success function) to generate a mathematical assessment of expected outcome. Indeed, several of James' sabermetric models are exercises in managerial economics (productivity of team, value of players, etc.), and this focus eventually landed him a position as senior baseball operations advisor with the Boston Red Sox (Neyer 2002). The original formula (James, 1980) appears as follows:

$$\text{Expected wins}_{i,j} = \frac{rs_{i,j}^{\alpha}}{rs_{i,j}^{\alpha} + ra_{i,j}^{\alpha}} \qquad (1)$$

where $rs_{i,j}$ represents runs scored by team i in season j, $ra_{i,j}$ represents runs allowed by team i in season j, and the original model restricts the noise parameter α to the value 2. The name of the model was derived from this restriction, though the model's association with the Pythagorean Theorem in geometry is quantitative rather than fundamental. The noise parametric restriction imposed by James (1980) is interesting for a couple of reasons. In the case of major league baseball, this restricted model has subsequently been shown to be a close approximation to the best fit noise parameter and by far the best integer choice. From a game-theoretic perspective, the expected wins percentage formula in (1) is an additive or Tullock-form contest success function (CSF) with noise parameter equal to 2 (Ehrlich et al., 2020). Within a symmetric contest model, such a CSF governs the "least noisy contest" (possesses the highest noise parameter value) of all such contests that generate a pure strategy Nash equilibrium.

Indeed, sports seasons can be classified as proportional prize labor contests, where downstream contest inputs (i.e., runs or points rather than physical efforts) are observed[1]. The literature on labor contests is substantial (for recent contributions, see, e.g., Jasina and Rotthoff 2012, Shaffer 2006, Ridlon and Shin 2013, or Gopalakrishna et al. 2016). Sports seasons are a transparent laboratory within which to decompose realized outcome into components of expected outcome and chance. Contest success functions have a rich history in the economics, marketing, management, and sports literatures in describing rent-seeking (lobby), firm strategy, firm research and development, sports matches, conflict, litigation, and other activities. Though referred to as a Tullock-form CSF for its use in rent-seeking contests (Tullock 1980), the additive-form CSF was first applied to the economics of firm advertising (Friedman 1958; Mills 1961) and later to the economics of sporting contests (El-Hodiri and Quirk 1971)[2]. Bill James presented the Pythagorean wins expectation model in the same year that Tullock presented his contest model of rent-seeking. James has stated that his economic education was important to his development of sabermetric models. He developed the Pythagorean Expected Wins formula through trial and error and did not estimate the exponential parameter. Subsequently, there has been an extended effort to refine his restricted model for baseball and other settings (see, e.g., Horowitz 1994; Davenport and Woolner, 1999; Miller 2007; Tung 2008; Cochran and Blackstock 2009; Winston 2012; Ehrlich et al., 2019).

In the present paper, we apply two alternative models from contest theory—the serial and difference-form contest success functions—to the problem of output (wins) estimation. We then transform each function to achieve linearity in the (CSF noise) parameter and estimate the parameter utilizing Major League Baseball team win data (from 2003-2015). We find the serial CSF estimator explains more variation (i.e., has a higher $R^2$) than the traditional Tullock-form model (or the difference-form model). Relative to the improvements achieved by correcting the parameter in James' original model, the efficiency gain from serial CSF estimation is dramatic. We obtain consistent results when we apply the least squares regression as well as the cross-validation approaches. We discuss differences in the forms of the respective CSFs in our discussion of these results. The paper constitutes a test of alternative CSF forms within a real-world labor contest. Our study also builds upon prior work finding the difference form CSF outperforms the traditional Pythagorean model in a simulation (Ehrlich et al., 2020). Similar to their study, we find the difference form CSF outperforms the original Pythagorean model by James (1980). Yet, we find the difference form CSF does not outperform the optimized Tullock-form model. Moreover, we extend this research by illustrating how the serial CSF outperforms both models, where lower Mean Square Error (MSE) and higher coefficient of determination ($R^2$) indicate better model performance.

## 2   The General Pythagorean Expected Wins Model

In its general form, the Pythagorean Expected Wins Model relaxes the assumption that α=2 and appears as in equation (1). Many follow-up studies (e.g., Davenport and Woolner, 1999; Baseball Reference; Heipp, 2004) have examined the accuracy of α=2, with the result that from year-to-year, the most accurate exponent ranges from 1.75 to 2.05 (Rothman, 2014). Here α(> 0) is defined in contest theory as the "noise" or level of determinism in the mapping from inputs (runs scored and allowed) to output (wins) (see, e.g., Nti 1999). Regardless of relative input levels (>0), expected win proportion goes to 0.5 when α = 0. As a result, as α → 0, the mapping becomes very noisy. For α close to 0 (e.g., for α = .0001), a team that scores 1,000 runs and allows only 1 run over the course of a season is expected to win approximately half of its games. Such a number of wins is not possible, however, as the minimum number of runs needed to win 81 games is 81. For baseball, then, we do not expect α to be zero. Alternatively, we do not expect α to be "too large" in baseball. For α = 1000, consider a team that scores 980 runs and allows 1000. Such a team has a wins expectation of approximately 0. A team with a net runs deficit of 20 cannot lose all of its games, however. Then, we expect α to be "somewhere in the middle" and perhaps very close to James' original specification.

Before estimating the noise parameter, α, let us transform the standard Pythagorean model such that the right hand side is linear in the parameter. Such a transformation allows us to estimate α using a linear (OLS) regression model.

$$\ln\left(\frac{1}{\text{EWP}_{i,j}} - 1\right) = \alpha \cdot \ln\left(\frac{\text{ra}_{i,j}}{\text{rs}_{i,j}}\right) \qquad (2)$$

---

[1] One can view a team's season record as a proportional prize contest between that team and its set of opponents. For example, a team that finishes a season with a record of 30 wins and 20 losses has won a 0.6 share of matches against opponents (has lost a 0.4 share of matches to opponents).

[2] El-Hodiri was an economics professor at the University of Kansas at the same time that James was an undergraduate economics student there. According to a personal correspondence from James, James and El-Hodiri never crossed paths. Their pioneering uses of CSFs to describe sport outcomes were independently conceived.

We also consider two alternative contest success functional forms: the difference form and the serial form. These respective forms appear as follows.

$$\text{Difference form: } EWP_{i,j} = \frac{1}{1+e^{\alpha(ra_{i,j}-rs_{i,j})}} \tag{3a}$$

$$\text{Serial Form: } \min\left(EWP_{i,j}, (1-EWP_{i,j})\right) = \frac{1}{2} \cdot \left(\frac{\min(rs_{i,j},\ ra_{i,j})}{\max(rs_{i,j},\ ra_{i,j})}\right)^{\alpha} \tag{3b}$$

The serial form, an obscure CSF form, does not directly estimate win proportion. Rather, it estimates, for each observation, the minimum of win proportion and its probabilistic complement (Corchon and Dahm, 2010). In discovering the serial CSF form, Alcalde and Dahm (2007) write: "We introduce the serial contest by building on the desirable properties of two prominent contest games. This family of contest games relies both on relative efforts (as Tullock's proposal) and on absolute effort differences (as difference-form contests)." As the serial form is an anonymous measure (such that a permutation of arguments generates a corresponding permutation of win probabilities), win proportion estimates are easily obtained for each observation. The serial form is an obscure CSF but is a hero in the present story. Similar transformations of the difference form and serial form CSFs give us the following linear models.

$$\text{Difference Form: } \ln\left(\frac{1}{EWP_{i,j}}-1\right) = \alpha(ra_{i,j}-rs_{i,j}) \tag{4a}$$

$$\text{Serial Form: } \ln\left(2 \cdot \min\left(EWP_{i,j}, (1-EWP_{i,j})\right)\right) = \alpha \cdot \ln\left(\frac{\min(rs_{i,j},\ ra_{i,j})}{\max(rs_{i,j},\ ra_{i,j})}\right) \tag{4b}$$

It is interesting to note that the difference form model is equivalent to a log-linear form of the model proposed by Jones and Tappin (2005). The serial model is most distinct among the models considered in terms of functional form. By taking advantage of the anonymity property of standard CSF models, it is able to employ a sparse ratio form without an additive restriction in the denominator.

## 3  Data

In order to estimate the models discussed above, we find thirteen years (2003-2015) of season win proportions, season runs scored, and season runs allowed for each Major League Baseball (MLB) team at ESPN. As such, we obtain 390 team-season observations and use this data to estimate the proposed models. Table 1 reports summary statistics.

**Table 1**: Summary Statistics

| Variable | Obs | Mean | Std. Dev. | Min | Max |
| --- | --- | --- | --- | --- | --- |
| Wins | 390 | 80.98462 | 11.14682 | 43 | 105 |
| Losses | 390 | 80.98205 | 11.11483 | 57 | 119 |
| Runs Scored | 390 | 729.2385 | 81.46349 | 513 | 968 |
| Runs Allowed | 390 | 729.2385 | 85.90975 | 525 | 971 |

## 4 Results and Discussion

We use the ordinary least square (OLS) regression method for the model specified in 2, 4a, and 4b. Table 2 reports the results.

**Table 2:** CSF Estimation Results

|  | Original Tullock model (James 1980); α = 2 | Tullock model with OLS-estimated α | Diff. model with OLS-estimated α | Serial model with OLS-estimated α |
|---|---|---|---|---|
| α | 2 | 1.859 | 0.00254 | 1.032 |
| t-stat: |  |  |  |  |
| H0 = 0 | -- | 50.07*** | 49.88*** | 52.56*** |
| H0 = 2 | -- | 3.79*** | 38,729*** | 49.31*** |
| $R^2$ | -- | 0.866 | 0.865 | 0.877 |
| RMSE | 4.069 | 4.0722 | 4.081 | 3.6836 |
| n | 390 | 390 | 390 | 390 |

Note. T-stat tests the hypothesis that the estimated parameter, α, is statistically and significantly different from 0 in the first row and 2—the value of James' original parameter—in the second row. * $p < 0.10$; ** $p < 0.05$; *** $p < 0.01$.

The findings in Table 2 are important for two reasons. First, the results indicate the serial model outperforms both the original Tullock model (James 1980) and the optimized Tullock and difference-form models. The serial model has the smallest mean squared error (MSE) and the largest $R^2$. Second, the estimated parameters for the optimized Tullock and difference-form models are similar to previous estimates. Using simulated data, Ehrlich et al. (2020) estimated the parameter, α, to be 1.722 and 0.003 for the Tullock and difference-form models, respectively. We estimate α to be 1.859 and 0.00254 for the same models.

Across the three models, then, the difference or ratio between runs scored and runs allowed is a highly explanatory variable in general. As the models are transformed, we must modify the residuals so that they are posed in probability terms. Root mean squared error values suggest that estimates of team win proportion are typically within approximately .025 units of the true win proportion, on average. This is equivalent to approximately 4 wins over a 162-game season $((.025 \cdot 162) \approx 4)$. Thus, the Pythagorean expected wins model possesses considerable explanatory power and efficiency for a single-parameter (simple CSF) model.

More specifically, the results indicate that the serial model improves considerably upon the historically used Tullock-form CSF model in terms of $R^2$ and root-MSE. Changing from James' restricted model to an optimized serial CSF achieves five times the root-MSE reduction than does changing from James' restricted model to an optimized Tullock-form model. That is to say, the improvement achieved from replacing James' initial parametric value guess with an OLS-estimated value is dwarfed by the improvement from the selection of an alternative model in this case. In Table 3 below, root-MSEs are represented in terms of wins per 162 games.

The serial model achieves a large reduction in root-MSE. Indeed, its estimation performance is considerably different from each of the alternative models employed. Among OLS-estimated models, the difference-form model performs worst among optimized models but still better than James' restricted Tullock model. This finding builds upon earlier work by Ehrlich et al. (2020) comparing the difference-form CSF and the Tullock-form CSF. However, our findings illustrate the serial model outperforms both models compared by Ehrlich et al. (2020).

## 5 Cross-Validation Approach

In addition to the individual analyses of the three models presented above, we apply a cross-validation technique to compare the predictive accuracy of these models. Cross-validation is a model validation technique for assessing how the results of a statistical analysis generalize, widely used for its simplicity and apparent universality (Arlot & Celisse, 2010). Because of the popularity of the method in comparing the performance of different modelling specifications, this technique is ideally suitable to compare the three models in our win prediction models.

There are multiple approaches to cross-validation. We use the leave-one-out approach, a particular case of the leave-p-out cross-validation. This method splits the data into two parts: training and validation sets. In each set, one of the 390 observations is considered to be a validation set and the rest of the remaining 389 observations are considered to be training set. This process is repeated for all other points in the dataset. In this process, the validation error is calculated in each step. At the end, we calculate the Mean Square Error (MSE) as the average of the errors, defined as:

$$LOOCV_{390} = \frac{1}{390} \sum_{i=1}^{390} MSE_i \qquad (5)$$

We repeat this process for each of the three models. The results are reported in Table 3.

**Table 3:** Leave-One-Out Cross-Validation Results

| Accuracy Method | Model | | |
|---|---|---|---|
| | Estimated Tullock Model | Difference Model | Serial Model |
| Root MSE | 4.0836 | 4.0926 | 3.6964 |
| Mean Absolute Errors (MAE) | 3.2807 | 3.2932 | 2.9357 |

The results are consistent with the earlier models. In addition to the RMSE discussed earlier, the cross-validation results show Mean Absolute Errors (MAE) which is a conceptually simpler and more interpretable than RMSE because it does not require the use of squares or square roots. In simple terms, we would interpret MAE as the average absolute vertical distance between each point in a scatter plot. Overall, the technique allows us to see the differences among the models. The results confirm that the serial model still turns out to be the best compared to the Tullock model and the difference model.

# 6  Conclusion

The present work demonstrates that insights from mathematical applications and contest theory toward better team quality estimation and prediction. Such knowledge is valuable, as future outcomes depend upon expected rather than realized outcomes. From a sports analytics perspective, it is important to understand how reproducible a given result may be in the next time period. In the setting of Major League Baseball (2003-2015), the serial CSF, though an obscure CSF form, leads to a dramatic reduction in root-MSE as compared to more prominent (and traditionally used) CSF forms such as the original Pythagorean expected wins percentage model. By taking advantage of the anonymity property of standard CSF models, it is able to employ a sparse ratio form without an additive term in the denominator. These results are consistent across multiple models as shown by the leave-one-out cross-validation approach. A future work might consider how the alternative models considered herein compare in alternative sports leagues.

# References

Alcalde, José, and Matthias Dahm. "Tullock and Hirshleifer: a meeting of the minds." *Review of Economic Design* 11, no. 2 (2007): 101-124.

Anthony, Tyler, Tim Kahn, Briana Madison, Rodney J. Paul, and Andrew Weinbach. "Similarities in fan preferences for minor-league baseball across the American Southeast." *Journal of Economics and Finance* 38, no. 1 (2014): 150-163.

Arlot, Sylvain, and Alain Celisse. "A survey of cross-validation procedures for model selection." *Statistics surveys* 4 (2010): 40-79.

Pythagorean Theorem of Baseball - BR Bullpen. Accessed October 19, 2020. https://bit.ly/3bS119m

Cochran, James J., and Rob Blackstock. "Pythagoras and the national hockey league." *Journal of Quantitative Analysis in Sports* 5, no. 2 (2009).

Corchón, Luis, and Matthias Dahm. "Foundations for contest success functions." *Economic Theory* 43, no. 1 (2010): 81-98.

Davenport, Clay, and Keith Woolner. "Revisiting the pythagorean theorem:Putting Bill James' Pythagorean Theorem to the test." *Baseball Prospectus* (1999). Accessed October 19, 2020. https://bit.ly/3vHaVmq

Drayer, Joris, Daniel A. Rascher, and Chad D. McEvoy. "An examination of underlying consumer demand and sport pricing using secondary market data." *Sport Management Review* 15, no. 4 (2012): 448-460.

ESPN (2019). "ESPN Enterprises". Accessed (n.d.), 2020. https://www.espn.com/mlb/standings

Ehrlich, Justin Andrew, Christopher Boudreaux, James Boudreau, and Shane Sanders. "Estimating Major League Baseball Team Quality through Simulation: An Analysis of an Alternative Pythagorean Expected Wins Model." *Mathematics and Sports* 1, no. 1 (2020).

Ehrlich, Justin, Shane Sanders, and Christopher J. Boudreaux. "The relative wages of offense and defense in the NBA: a setting for win-maximization arbitrage?." *Journal of Quantitative Analysis in Sports* 15, no. 3 (2019): 213-224.

El-Hodiri, Mohamed, and James Quirk. "An economic model of a professional sports league." *Journal of political economy* 79, no. 6 (1971): 1302-1319.

Friedman, Lawrence. "Game-theory models in the allocation of advertising expenditures." *Operations research* 6, no. 5 (1958): 699-709.

Gopalakrishna, Srinath, Jason Garrett, Murali K. Mantrala, and Shrihari Sridhar. "Assessing sales contest effectiveness: the role of salesperson and sales district characteristics." *Marketing Letters* 27, no. 3 (2016): 589-602.

Hadley, Lawrence, Marc Poitras, John Ruggiero, and Scott Knowles. "Performance evaluation of national football league teams." *Managerial and Decision Economics* 21, no. 2 (2000): 63-70.

Heipp, Brandon. "W% estimators." (2004). Accessed October 19, 2020. http://gosu02.tripod.com/id69.html

Horowitz, Ira. "Pythagoras, Tommy Lasorda, and me: On evaluating baseball managers." *Social Science Quarterly* (1994): 187-194.

James, Bill. "Baseball Abstract." Manuscript: Lawrence, KS. (1980).

Jasina, John, and Kurt Rotthoff. "A model of promotion and relegation in league sports." *Journal of Economics and Finance* 36, no. 2 (2012): 303-318.

Miller, Steven J. "A derivation of the Pythagorean Won-Loss Formula in baseball." *Chance* 20, no. 1 (2007): 40-48.

Mills, Harlan.D. A study in promotional competition, in: F.M. Bass et al. (eds.), Mathematical Models and Methods in Marketing, R.D. Irwin, Homewood, (1961): 245-301. Reprinted from: Research Paper No. 101-103, December 1959, Mathematica, Princeton, N.J.

Mongeon, Kevin, and Jason Winfree. "Comparison of television and gate demand in the National Basketball Association." *Sport Management Review* 15, no. 1 (2012): 72-79.

Neyer, Rob. "Red Sox Hire James in Advisory Capacity". (2002). Accessed October 16, 2020. https://bit.ly/2NqkGEe

Nti, Kofi O. "Rent-seeking with asymmetric valuations." *Public Choice* 98, no. 3 (1999): 415-430.

Ridlon, Robert, and Jiwoong Shin. "Favoring the winner or loser in repeated contests." *Marketing Science* 32, no. 5 (2013): 768-785.

Rothman, Stanley. "A New Formula to Predict a Team's Winning Percentage." *The Baseball Research Journal, 43*(2) (2014): 97-105.

Shaffer, Sherrill. "War, labor tournaments, and contest payoffs." *Economics Letters* 92, no. 2 (2006): 250-255.

Tullock, Gordon, James M. Buchanan, and Robert D. Tollison. "Toward a Theory of the Rent-seeking Society." Efficient rent seeking 97 (1980): 112.

Tung, David. D. "Confidence Intervals for the Pythagorean Formula in Baseball." Available at: http://www.rxiv.org/pdf/1005.0020v1.pdf

Van Leeuwen, Linda, Shayne Quick, and Kerry Daniel. "The sport spectator satisfaction model: A conceptual framework for understanding the satisfaction of spectators." *Sport Management Review* 5, no. 2 (2002): 99-128.

Winston, Wayne. L. "Mathletics: How gamblers, managers, and sports enthusiasts use mathematics in baseball, basketball, and football". Princeton University Press (2012).

APPENDIX

**To go from (1) to (2):**

$$\text{EWP}_{i,j} = \frac{\text{rs}_{i,j}^{\alpha}}{\text{rs}_{i,j}^{\alpha} + \text{ra}_{i,j}^{\alpha}} \tag{A1}$$

$$\frac{1}{\text{EWP}_{i,j}} = \frac{\text{rs}_{i,j}^{\alpha} + \text{ra}_{i,j}^{\alpha}}{\text{rs}_{i,j}^{\alpha}} = 1 + \frac{\text{ra}_{i,j}^{\alpha}}{\text{rs}_{i,j}^{\alpha}} \tag{A2}$$

$$\left(\frac{1}{\text{EWP}_{i,j}} - 1\right) = \frac{\text{ra}_{i,j}^{\alpha}}{\text{rs}_{i,j}^{\alpha}} = \left(\frac{\text{ra}_{i,j}}{\text{rs}_{i,j}}\right)^{\alpha} \tag{A3}$$

$$\ln\left(\frac{1}{\text{EWP}_{i,j}} - 1\right) = \alpha \cdot \ln\left(\frac{\text{ra}_{i,j}}{\text{rs}_{i,j}}\right) \tag{A4}$$

**To go from (2) to (3a):**

$$\ln\left(\frac{1}{\text{EWP}_{i,j}} - 1\right) = \alpha \cdot \ln\left(\frac{\text{ra}_{i,j}}{\text{rs}_{i,j}}\right) \tag{A5}$$

$$\left(\frac{1}{\text{EWP}_{i,j}} - 1\right) = e^{\alpha(\text{ra}_{i,j} - \text{rs}_{i,j})} \tag{A6}$$

$$\left(\frac{1}{\text{EWP}_{i,j}}\right) = 1 + e^{\alpha(\text{ra}_{i,j} - \text{rs}_{i,j})} \tag{A7}$$

$$\text{EWP}_{i,j} = \frac{1}{1 + e^{\alpha(\text{ra}_{i,j} - \text{rs}_{i,j})}} \tag{A8}$$